\documentclass[aps,graphicx,twocolumn,epsfig]{revtex4}

\usepackage{epsfig}
\usepackage{amsmath}

\newcommand{\ud}{\mathrm{d}}
\newcommand{\myvec}[1]{{\bf {#1}}} 
\newcommand{\ket}[1]{|{#1}\rangle}                     
\newcommand{\bra}[1]{\langle{#1}|} 
\newcommand{\braket}[2]{\langle{#1}|{#2}\rangle}

\begin{document}

\author{C. Trefzger$^{1}$}
\author{M. Alloing$^{1}$} 
\author{C. Menotti$^{2}$}
\author{F. Dubin$^{1}$}  
\author{M. Lewenstein$^{1,3}$} 
\affiliation{$^1$ ICFO - Institut de Ciencies Fotoniques, Mediterranean Technology Park, 08860 Castelldefels (Barcelona), Spain \\
$^2$ INO-CNR BEC Center and Dipartimento di Fisica, Universit\`a di Trento, 
38123 Povo, Italy\\ 
$^3$ ICREA - Instituci\`o Catalana de Recerca i Estudis Avan\c cats, Lluis Companys 23, E-08010 Barcelona, Spain
}

\title{Quantum magnetism and counterflow supersolidity of up-down bosonic dipoles}

\begin{abstract}
We study a gas of dipolar Bosons confined in a two-dimensional optical lattice. Dipoles are considered to point freely  in both up and down directions perpendicular to the lattice plane. This results in a nearest neighbor repulsive (attractive) interaction for aligned (anti-aligned) dipoles. We find regions of parameters where the ground state of the system exhibits insulating phases with ferromagnetic or anti-ferromagnetic ordering, as well as with rational values of the average magnetization. Evidence for the existence of a novel counterflow supersolid quantum phase is also presented.
\end{abstract}

\maketitle

\section{Introduction}

Recently, the physics of ultracold dipolar gases has received growing attention. Experimental and theoretical studies are motivated by the long-range and anisotropic dipole-dipole interaction which  introduce a rich variety of quantum phases \cite{Lahaye:09}. For instance, supersolid and checkerboard phases are predicted in the phase diagram of polarized dipolar Bosons confined in a two-dimensional (2D) optical lattice \cite{Goral:02}. In bilayer samples where dipole interactions can also be attractive more exotic quantum phases are accessible, for example a pair-supersolid \cite{BB:Trefzger02}. 

Experiments with ultracold atoms have now highlighted the role of dipole-dipole interactions. Notably, with $^{52}$Cr, control via Feshbach resonances allows one to efficiently reduce contact interactions and enter a regime where magnetic dipole-dipole interactions become dominant \cite{Lahaye:07}. Polar molecules also appear very promising towards implementation of degenerate dipolar gases (see \cite{Lahaye:09} and References therein), in particular since the demonstration of ultra-cold rubidium-potassium, and lithium-cesium
molecules prepared in their ground rotovibrational state \cite{Ni:08,BB:Weidemuller}. 

Note that in most cases, if not all cases considered so far, dipolar gases are polarized, i.e. magnetic or electric dipoles point in the same direction.
This, however, does not have to be the case always. A prominent example is that of exciton gases. Excitons, which are bound electron-hole pairs, obviously carry an electric dipole moment that can {\it a priori} attain quite arbitrary directions. Nevertheless, in indirect quantum wells, electrons and holes are confined in spatially separated regions such that electric dipoles can be aligned \cite{Butov} or even anti-aligned in type-II heterostructures \cite{Danan}. 

Another extreme case concerns molecules that follow the Hund's rule (a) \cite{BB:Hund}. For such molecules the electric dipole is either parallel or anti-parallel to the direction of the magnetic moment. If one takes thus a sample of such molecules and polarizes it magnetically, say in the up direction, one will obtain in this way a dipolar gas with certain fraction of electric dipoles pointing up, and the remaining fraction pointing down. This is the situation that we study in the present paper. In fact there was recently a spectacular progress in cooling and trapping of magnetically confined neutral OH molecules. This progress opens important perspectives to study novel quantum phases with ultracold dipoles \cite{BB:Lev01, BB:Lev02}. In its ro-vibronic ground state OH is indeed a Hund's case (a) molecule, and it features both an electric $d = 1.67 D$ and a magnetic dipole moment $\mu = 1.2\mu_{\scriptscriptstyle \mathrm{B}}$ (with $\mu_{\scriptscriptstyle \mathrm{B}}$ being the Bohr magneton), which can be aligned or anti-aligned independently. Using these two degrees of freedom, OH molecules may be confined in a 2D optical lattice, where the dipolar interactions lead to rich quantum phases, as we show in this work.

We study here a sample of bosonic dipoles confined in a 2D lattice. We consider that dipoles are free to point in both directions normal to the lattice plane. This constitutes a novel ingredient compared to our previous works \cite{Menotti:07,Trefzger:08}, and results in a nearest neighbor interaction either repulsive for aligned dipoles, or attractive for anti-aligned ones. 
We consider the case of dipolar interactions to be relatively weak compared to the contact interactions, and remarkably we find that 
the dipole-dipole interactions dominate the physics of the system.    
Using a mean-field Gutzwiller approach, we show that the system presents Mott-insulating phases with ferromagnetic or anti-ferromagnetic ordering, as well as with fractional values of the average magnetization, depending on the imbalance between the population of dipoles pointing in opposite directions. We also find evidences of a novel {\it counterflow supersolid} quantum phase. The latter exhibits broken translational symmetry, namely a modulation of the order parameter on a length scale larger than the one of the lattice spacing, analogously to supersolid phases.

\section{Hamiltonian of the system}

\begin{center}
\begin{figure}[t]
\begin{center}
\includegraphics[width=1\linewidth]{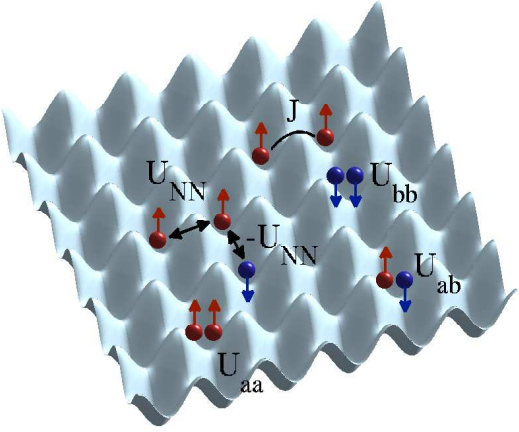}
\end{center}
\caption{Schematic representation of a 2D
optical lattice populated with dipolar Bosons polarized
in both directions perpendicular to the lattice plane. The particles feel repulsive
intra-species $U_{aa}$, $U_{bb}$, and inter-species $U_{ab}$ repulsive on-site energies.
The first nearest-neighbor interaction is repulsive $U_{\rm NN}>0$ for aligned dipoles, while it is attractive $-U_{\rm NN}$ for anti-aligned particles while the hopping term $J$ is equal for both the species.}
\label{FIG:UpDown}
\end{figure}
\end{center}

We consider a sample of dipoles in the presence of a 2D optical lattice, and an extra
confinement in the perpendicular direction. The dipoles are free to point in both directions normal to the lattice plane, as shown in Fig.~\ref{FIG:UpDown}. The system is described by the Hamiltonian 
\begin{eqnarray}
\hat{H} &=& \sum_{i,\sigma} \left[ \frac{U_\sigma}{2} \hat{n}_i^\sigma (\hat{n}_i^\sigma - 1) +  \frac{U_{\sigma\sigma^\prime}}{2}\hat{n}_i^\sigma \hat{n}_i^{\sigma^\prime} -  \mu_\sigma \hat{n}_i^\sigma \right]\nonumber \\
          &+& \frac{1}{2}\sum_{ i\neq j,\sigma}\frac{U_{NN}}{|r_{ij}|^3} \left[ \hat{n}_i^\sigma \hat{n}_j^\sigma - \hat{n}_i^\sigma \hat{n}_j^{\sigma^\prime}\right]  - J\sum_{\langle ij \rangle} [\hat{a}_i^\dag \hat{a}_j + \hat{b}_i^\dag \hat{b}_j ],\nonumber\\ 
          \label{EQ:H3D}
\end{eqnarray}
where  $\sigma = \big(a, b\big)$ indicates the type of species, i.e. dipoles pointing in the up and
down direction perpendicular to the 2D plane of the lattice, respectively. 
$U_{aa}$ and $U_{bb}$ are the on-site
energies for particles of the same species, while $U_{ab}$ is the on-site energy for different species. 
The long-range dipolar interaction potential decays as the inverse cubic power of the relative distance $r_{ij}$, which we
express it in units of the lattice spacing. For computational reasons, in most theoretical approaches the range is cutoff at a certain neighbors.
In the present work, we consider a range of interaction up to the 4th nearest neighbor.
The first nearest neighbor dipolar interaction is attractive (repulsive) for particles of the same
(different) species, with strength $U_{NN}>0$ (or respectively $-U_{NN}$), 
and we consider an equal tunneling coefficient $(J)$ for both the species, while
the densities of the dipoles pointing upwards and downwards are fixed by the corresponding chemical potentials $\mu_\sigma$. 
The on-site interactions have two contributions: one is arising from the s-wave scattering length
$U_s = \frac{4\pi\hbar^2 a_s}{m}\int \ud^3r \rho(\myvec{r})^2$, and the second one is due to the on-site dipole-dipole interaction
$U_{\rm dd} = \frac{1}{(2\pi)^3} \int \ud^3k \, \widetilde{U_{\rm dd}}(\myvec{k}) \;\widetilde{\rho}^{\;2}(\myvec{k})$, where $\widetilde{U_{\rm dd}}(\myvec{k})$
and $\widetilde{\rho}^{\;2}(\myvec{k})$ are the Fourier transform of the dipole potential and the density respectively. We consider that the s-wave scattering length is independent on the orientation of the dipoles. Instead, the on-site
dipolar contribution $U_{\rm dd}$ depends both on the orientation of the dipoles and on the geometry of the trapping potential, and it can be varied by changing the 
ratio between the vertical to the axial confinement. 
For simplicity we will focus on the specific case of a spherically symmetric confinement, where the on-site dipolar interactions
average out to zero $U_{\rm dd} = 0$, and the resulting on-site interactions are all equal to $U$. 
We consider the case of dipole-dipole interactions to be 600 times weaker with respect to the on-site
interaction, i.e. $U_{\rm NN} = U/600$ \cite{BB:Foote01}.

\subsection{Filling factor and imbalance}
\label{SEC:FillingImbalance}
The properties of the system are conveniently extracted using the operators given by the sum (filling factor) and by the difference (imbalance) of the two species number operators at each site of the lattice, namely by
\begin{equation}
\label{EQ:PMOperators}
\begin{split}
\hat{\nu}_i &= \frac{\hat{n}_i^a + \hat{n}_i^b}{2} \\
\hat{m}_i &= \frac{\hat{n}_i^a - \hat{n}_i^b}{2},
\end{split}
\end{equation}
which are simultaneously diagonal on a given Fock state $\ket{\nu,m}_i$.
Notice that the eigenvalues of these two operators are not independent. In fact, by fixing $\nu$ the eigenvalues of $\hat{m}_i$ can only assume 
$2\nu + 1$ values given by $m = \{-\nu,  -\nu+1,..., +\nu\}$, in complete analogy with the angular momentum operator $\hat{S}_i^2$ and its projection along the 
$z$ axis $\hat{S}_i^z$, as we will discuss in Sec. \ref{SEC:Low-Energy}. It is also useful to introduce the average magnetization of the system, defined as
\begin{equation}
M = \frac{1}{N_{\scriptscriptstyle \mathrm{S}}}\sum_i m_i,
\end{equation}
where $N_{\scriptscriptstyle \mathrm{S}}$ is the total number of lattice sites.
 
\begin{center}
\begin{figure}[h!]
\begin{center}
\includegraphics[width=1\linewidth]{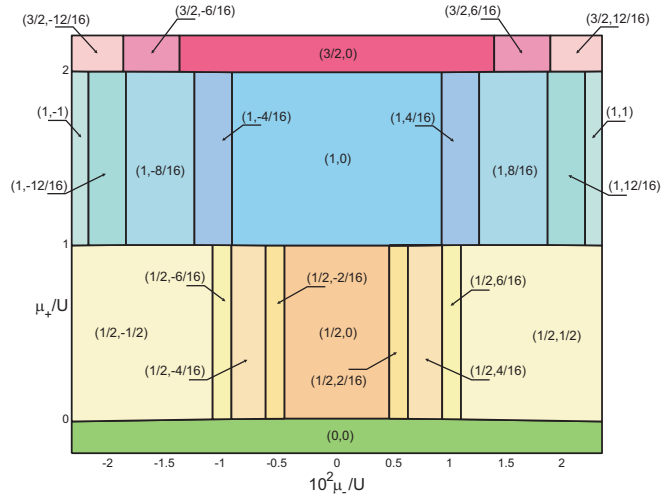}
\end{center}
\caption{Ground state of the system at $J=0$, calculated for a $4\times 4$ elementary cell satisfying periodic boundary conditions, and 
$U_{\rm NN} = U/600$. The text in parentheses $(\nu,M)$ indicates the filling factor $\nu$ and the average magnetization $M$.}
\label{FIG:GSJ0}
\end{figure}
\end{center}

Substituting Eqs. (\ref{EQ:PMOperators}) into equation (\ref{EQ:H3D}) allows us to express the system Hamiltonian as $\hat{H} = \hat{H}_{\rm 0}^\nu + \hat{H}_{\rm 0}^m + \hat{H}_{\rm 1}^{\nu m}$,
where 
\begin{eqnarray}
\hat{H}_{\rm 0}^\nu       &=& \sum_i \Big[-2\mu_{\scriptscriptstyle \mathrm{+}} \hat{\nu}_i + 2U\hat{\nu}_i\Big(\hat{\nu}_i - \frac{1}{2}\Big)\Big]\label{EQ:HNUM01} \\
\hat{H}_{\rm 0}^m         &=&\sum_i \Big[-2\mu_{\scriptscriptstyle \mathrm{-}} \hat{m}_i + 2U_{NN}\sum_{j\neq i}\frac{\hat{m}_i\hat{m}_j}{|r_{ij}|^3} \Big]\label{EQ:HNUM02} \\
\hat{H}_{\rm 1}^{\nu m} &=& - J\sum_{\langle ij \rangle} [\hat{a}_i^\dag \hat{a}_j + \hat{b}_i^\dag \hat{b}_j ]. \label{EQ:HNUM03}
\end{eqnarray}
In Eqs. (\ref{EQ:HNUM01},\ref{EQ:HNUM02}), we have introduced the chemical potentials
\begin{equation}
\mu_\pm = \frac{\mu_a \pm \mu_b}{2},
\end{equation}
which respectively fix the eigenvalues of the filling factor ($+$), and the imbalance operators ($-$) in Eq. (\ref{EQ:PMOperators}).
In the following we consider $\hat{H}_{\rm 1}^{\nu m}$ to be a small perturbation on the interaction terms (\ref{EQ:HNUM01}) and (\ref{EQ:HNUM02}).
In the limit where $U \gg \big(U_{NN}$,$J\big)$, the ground state of the system is found to be a uniform distribution of constant filling factor 
$\nu_i = \bar{\nu}$ at each site of the lattice. The value of $\bar{\nu}$ is fixed by $\mu_{\scriptscriptstyle \mathrm{+}}$, and 
can be integer as well as semi-integer. 
This is better understood
at $J=0$, where we can calculate the expectation value of 
$\hat{H}_{\rm 0}^\nu$ on a given classical distribution of atoms in the lattice $\ket{\Phi} = \prod_i \ket{\nu_i,m_i}_i$, as follows
\begin{equation}
\label{EQ:EnergyNu}
\bra{\Phi} \hat{H}_{\rm 0}^\nu \ket{\Phi} = \sum_i \Big[-2\mu_{\scriptscriptstyle \mathrm{+}} \nu_i + 2U\nu_i\Big(\nu_i - \frac{1}{2}\Big)\Big],
\end{equation}
where $\nu_i = \bra{\Phi} \hat{\nu}_i \ket{\Phi}$. In the right hand-side of Eq. (\ref{EQ:EnergyNu}) all sites $i$ are equal, and like in the homogeneous case of a 
Bose-Hubbard Hamiltonian at $J=0$ \cite{BB:Fisher}, the minimum of Eq. (\ref{EQ:EnergyNu}) is provided by a uniform distribution in the lattice, where $\nu_i = \bar{\nu}$ at each site.
Instead, for a given $\bar{\nu}$, finding the distribution of $m_i = \bra{\Phi} \hat{m}_i \ket{\Phi}$, which minimize the expectation value
\begin{equation}
\label{EQ:EnergyM}
\bra{\Phi} \hat{H}_{\rm 0}^m \ket{\Phi} = \sum_i \Big[-2\mu_{\scriptscriptstyle \mathrm{-}} m_i + 2U_{NN}\sum_{j \neq i} \frac{m_i m_j}{|r_{ij}|^3} \Big],
\end{equation}
is non-trivial due to the presence of the long-range interaction. However, we can qualitatively argue that for 
$|\mu_{\scriptscriptstyle \mathrm{-}}| \gg U_{NN}$, the minimum of the energy (\ref{EQ:EnergyM}) is obtained for 
$m_i = \nu \times{\rm sign} \big[\mu_{\scriptscriptstyle \mathrm{-}}\big]$ $\forall i$, which corresponds to a ferromagnetic phase (FM) of average magnetization 
$M=\nu \times{\rm sign} \big[\mu_{\scriptscriptstyle \mathrm{-}}\big]$, where only particles of one species are present. Instead, for $\mu_{\scriptscriptstyle \mathrm{-}} = 0$,
a succession of nearest neighbors with $m_i = \nu$ and $m_j = -\nu$ provides the minimum of Eq. (\ref{EQ:EnergyM}), and the phase is anti-ferromagnetic
 (AM), i.e. $M=0$. In other words, the spatial distribution of the particles is given by sites occupied from the species $a$ alternated 
with sites occupied by the species $b$ in a checkerboard-like structure.
In Fig. \ref{FIG:GSJ0} we plot the ground state phase diagram at $J=0$, in the $\mu_{\scriptscriptstyle \mathrm{-}}$ vs. $\mu_{\scriptscriptstyle \mathrm{+}}$ plane,
where the text in parentheses indicates filling factor and average magnetization, i.e. $\nu$ and $M$ respectively. Note that between the AM, and the FM phases, we find 
magnetic phases which present non trivial rational values of the average magnetization (RM). The precise choice of the cutoff range in the dipolar interaction, and the lattice size, determine the fractional character of the allowed ground state magnetization.
In the next section, we include the presence of tunneling. The physical properties of the system are described by an effective Hamiltonian, supported by the existence of a low-energy subspace. Hence, the theoretical description of the system cannot be based
on standard mean-field theory, which is not suitable to correctly describe the ground state of the system, as we discuss in the following section.

\subsection{Low-energy subspace and effective Hamiltonian}
\label{SEC:Low-Energy}
\begin{center}
\begin{figure}[h!]
\begin{center}
\includegraphics[width=0.85\linewidth]{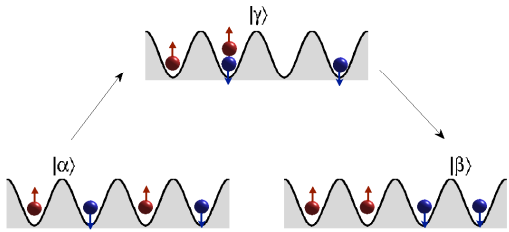}
\end{center}
\caption{Schematic representation of a two-particle hopping between the states $\ket{\alpha}$ and $\ket{\beta}$, belonging to the low-energy subspace at 
$\nu = 1/2$. These states are coupled through virtual excitations to the states $\ket{\gamma}$ by single-particle hopping.}
\label{FIG:Subspace}
\end{figure}
\end{center}

The ground state of the system at $J=0$ is described by a product over single-site Fock states of the type
\begin{equation}
\label{EQ:Alphas01}
\ket{\alpha} = \prod_i \ket{\nu,m_i}_i,
\end{equation}
with uniform total on-site occupation $2\nu$. As single particle hopping changes the total on-site population, it breaks the translational invariance of the ground state with respect to the total on-site occupation $2\nu$. 
The energy cost of these excitations is of the order of the on-site interaction energy $U$, and is therefore very costly in the limit where $U \gg \big(U_{NN},J\big)$.
On the contrary, exchanging two particles from nearest neighboring sites does not require such a
large amount of energy. This defines a low-energy subspace spanned by the $\ket{\alpha}$ configurations at 
constant filling factor $\nu$ of Eq. (\ref{EQ:Alphas01}), which is energetically well separated from the rest of the Hilbert space in the limit of parameters we consider. 
Thus, a successful description of such a system is obtained through an effective Hamiltonian $\hat{H}_{\rm eff}$ restricted to the low-energy subspace, where
single-particle hopping is suppressed and tunneling is included at second order in perturbation theory. The validity of the effective Hamiltonian relies on the 
existence of this low-energy subspace well separated in energy from the subspace of virtual
excitations, to which it is coupled via single-particle hopping. The relevant virtual subspace is then obtained from the states $\ket{\alpha}$ via single particle hopping, and it is spanned by the states
\begin{equation}
\label{EQ:SubspaceCSS}
\begin{split}
\ket{\gamma_{ij}^{(a)}} &= \frac{\hat{a}_i^\dag \hat{a}_j }{\sqrt{n_j^a(n_i^a + 1 )}} \ket{\alpha} \\
\ket{\gamma_{ij}^{(b)}} &= \frac{\hat{b}_i^\dag \hat{b}_j }{\sqrt{n_j^b(n_i^b + 1 )}} \ket{\alpha},
\end{split}
\end{equation}  
as schematically represented in Fig. \ref{FIG:Subspace}. 
This situation is in fact similar to the one discussed in \cite{BB:Trefzger02} for a bilayer optical lattice, and therefore we apply 
the same technique to compute $\hat{H}_{\rm eff}$. In the basis of constant total on-site population $2\nu$, the matrix elements of such a Hamiltonian at second order in perturbation theory are given by
\begin{eqnarray}
\bra{\alpha}\hat{H}_{\rm eff}\ket{\beta} &=& \bra{\alpha}\hat{H}_{\rm 0} \ket{\beta} - \frac{1}{2} \sum_\gamma \bra{\alpha} \hat{H}_{\rm 1}^{\nu m} \ket{\gamma} \bra{\gamma}\hat{H}_{\rm 1}^{\nu m} \ket{\alpha}\nonumber \\
& & \times \left[ \frac{1}{E_\gamma - E_\alpha} + \frac{1}{E_\gamma - E_\beta} \right] 
\end{eqnarray}
where $\hat{H}_{\rm 0} = \hat{H}_{\rm 0}^\nu + \hat{H}_{\rm 0}^m$, given by the sum of the interaction terms (\ref{EQ:HNUM01}) and (\ref{EQ:HNUM02}), is diagonal on the states $\ket{\alpha}$, and the
single-particle tunneling term $\hat{H}_{\rm 1}^{\nu m}$ of Eq. (\ref{EQ:HNUM03}) is treated at second order.
For a given state $|\alpha\rangle$,
\begin{eqnarray}
\label{EQ:Den02}
E_{\gamma_{ij}}-E_\alpha = U + U_{NN} \Delta m^{ij}_{\rm NN},
\end{eqnarray}
with $\Delta m^{ij}_{\rm NN}=\sum_{k\neq i}2m_k/|r_{ik}|^3 - \sum_{k\neq j} 2m_{k}/|r_{jk}|^3 - 1$, where $m_i$
indicates the population imbalance at site $i$ of Eq. (\ref{EQ:PMOperators}). For $U \gg U_{NN}$, the denominators
$E_{\gamma_{ij}}-E_\alpha$ are all of order $U$, which leads to
\begin{eqnarray}
\hat{H}_{\rm eff}^{(0)}  &=&   \hat{H}_{\rm 0}^\nu - \frac{2J^2}{U} \sum_{\langle ij \rangle} \hat{\nu}_i(\hat{\nu}_j+1) \nonumber \\
&+& \hat{H}_{\rm 0}^m - \frac{2J^2}{U} \sum_{\langle ij \rangle} \left[ \hat{m}_i \hat{m}_j + \hat{c}_i^\dag \hat{c}_j \right], \label{Heff0_num}
\end{eqnarray}
where $\hat{c}_i=\hat{a}_i \hat{b}_i^\dag$ and $\hat{c}_i^\dag = \hat{a}_i^\dag \hat{b}_i$ are composite operators, corresponding to the creation of a particle of one species and a hole of the other species, such that 
\begin{equation}
\label{EQ:CreationAnn}
\begin{split}
\hat{c}_i |\nu,m_i\rangle &= \sqrt{\nu(\nu+1) - m_i(m_i-1)}|\nu,m_i - 1\rangle\\
\hat{c}_i^\dag|\nu,m_i\rangle &= \sqrt{\nu(\nu+1) - m_i(m_i+1)}|\nu,m_i + 1\rangle,
\end{split}
\end{equation}
while their commutation relation obeys $[\hat{c}_i,\hat{c}_j^\dag] = -2\hat{m}_i\delta_{ij}$.

For a given $\nu$, the second line of the Hamiltonian ($\ref{Heff0_num}$) can be equivalently written in terms of the spin operators at site $i$ \cite{BB:Kuklov,BB:Altman}, given by
\begin{equation}
\label{EQ:Spin}
\hat{\myvec{S}}_i = \frac{1}{2} \sum_{\rm uu^\prime} \hat{a}_{\rm iu}^\dag \vec{\sigma}_{\rm uu^\prime}\hat{a}_{\rm iu^\prime},
\end{equation}
where $\vec{\sigma} = (\sigma_x,\sigma_y,\sigma_z)$ are the Pauli matrices, and $u = \big(a, b\big)$ indicates each species. Thus, the creation and annihilation operators
(\ref{EQ:CreationAnn}) become $\hat{c}_i^\dag = \hat{S}_i^x + i \hat{S}_i^y$ and $\hat{c}_i = \hat{S}_i^x - i \hat{S}_i^y$ respectively, while the imbalance operator 
is given by $\hat{m}_i = \hat{S}_i^z$ as already anticipated in Sec. \ref{SEC:FillingImbalance}. Therefore, in a spin representation as in Eq. (\ref{EQ:Spin}), the second line of the Hamiltonian ($\ref{Heff0_num}$) acts as a Heisenberg spin Hamiltonian (see e.g. \cite{BB:Sachdev}). The chemical potential $\mu_{\scriptscriptstyle \mathrm{-}}$ plays the role of an external magnetic field along the $z$ axis. The interplay between $\mu_{\scriptscriptstyle \mathrm{-}}$, and the long-range interaction
determines the magnetic ordering of the system, as we will discuss in the next section.

\section{Mean-field}
In this section, we provide a mean-field solution to the
effective Hamiltonian (\ref{Heff0_num}) to investigate quantum phases 
of the system. We identify the different phases  
through the composite order parameters $\langle \hat{c}_i \rangle$, as well as both the single-particle ones
$\langle \hat{a}_i \rangle$, and $\langle \hat{b}_i \rangle$.

For every subspace at constant filling, we find that the system 
presents three different kinds of phases. The {\it Mott-insulating} phase (MI), with a well
defined number of particles at each site of the lattice, 
and absence of any low-energy transport \cite{BB:Fisher}. 
The MI is characterized by vanishing $\langle \hat{c}_i \rangle = \langle \hat{a}_i \rangle = \langle \hat{b}_i \rangle = 0$,
and depending on the value of $\mu_{\scriptscriptstyle \mathrm{-}}$ presents FM, AM or RM ordering.  
The second quantum phase, a {\it super-counter-fluid} phase (SCF), exhibits on-site density fluctuations where a net transport of atoms is suppressed but a counterflow is present, for which 
the currents of the two species have equal amplitudes but opposite directions \cite{BB:Kuklov}.
In the SCF phase, while the single-particle order parameters still vanish $\langle \hat{a}_i \rangle = \langle \hat{b}_i \rangle = 0$,
the composite order parameters are non-zero $\langle \hat{c}_i \rangle \neq 0$, indicating the presence of counterflow.
We also find evidence of a third and novel quantum phase, namely a {\it counterflow supersolid} phase (CSS).
The CSS is characterized by vanishing single-particle order parameters $\langle \hat{a}_i \rangle = \langle \hat{b}_i \rangle = 0$,
and non-vanishing composite order parameters $\langle \hat{c}_i \rangle \neq 0$, coexisting with broken translational
symmetry, namely, a modulation of both $m_i$, and $\langle \hat{c}_i \rangle$ on a scale larger than the one of the lattice
spacing, analogously to the supersolid phase.

To determine the insulating phases we perform a perturbative treatment
at first order in the composite order parameters $\psi_i = \langle \hat{c}_i \rangle$, which 
allows us to compute the boundaries of the insulating lobes. Furthermore,
we solve the time dependent Gutzwiller equations in
imaginary time to determine the nature of the SCF-CSS
phases outside the lobes.

\subsection{Insulating lobes}
\begin{center}
\begin{figure}[h!]
\begin{center}
\includegraphics[width=1\linewidth]{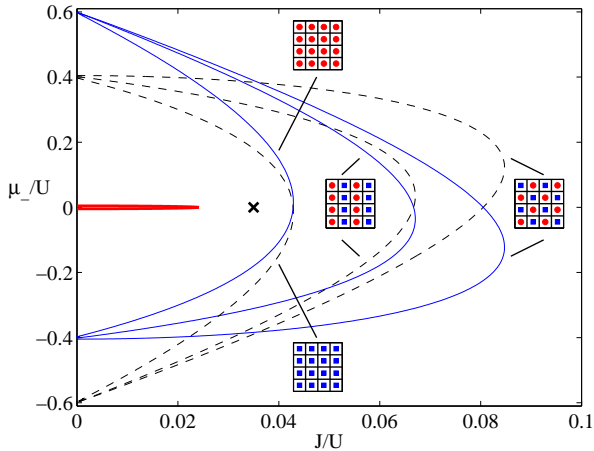}
\end{center}
\caption{Insulating lobes at $\nu=1/2$, for $\mu_{\scriptscriptstyle \mathrm{+}} = 0.4U$, and $U_{\rm NN} = U/600$.
The thick red line is the anti-ferromagnetic insulating state calculated with the effective Hamiltonian 
$\hat{H}_{\rm eff}^{(0)}$. The thin blue lines and the black dashed lines represent insulating lobes calculated with respect to single-particle-hole excitations of the species $a$ and species $b$ respectively, and are some of the many dominant configurations of the $J=0.035U$ and $\mu_{\scriptscriptstyle \mathrm{-}} = 0$ point (cross). The two species are sketched as a plain dots ($a$) and squares ($b$).}
\label{FIG:StabilityCSS}
\end{figure}
\end{center}

The low-energy subspace is spanned by the classical distribution of atoms 
in the lattice $\ket{\alpha}$ of Eq. (\ref{EQ:Alphas01}).
Similarly to the two layer system discussed in \cite{BB:Trefzger02}, in the limit where $U \gg U_{\rm NN}$, asymptotically all classical states
$\ket{\alpha}$ become stable with respect to single-particle-hole excitations
and develop an insulating lobe at finite $J$. The
energy of single particle-hole excitations is of the order of
$U$ at $J=0$ and is given by the width of the lobes at finite $J$
(see, e.g., the thin blue/black lobes in Fig. \ref{FIG:StabilityCSS}).

Instead, the low-lying excitations remain within the subspace and are obtained by adding (PH) or removing (HP) one composite, made of a particle of 
the upper-polarized dipoles (species $a$), and a hole of the lower-polarized dipoles (species $b$), at the $i$-th site of the lattice. This corresponds to flipping the direction of a dipole at the site $i$, respectively from down to up (PH) or from up to down (HP). For any given configuration $\ket{\alpha}$, one can calculate the corresponding energy costs using the diagonal terms of the effective Hamiltonian (\ref{Heff0_num}), which are respectively given by 
\begin{equation}
\begin{split}
\label{EQ:PHE_CSS}
E_{\scriptscriptstyle \mathrm{PH}}^i(J) &=-2\mu_{\scriptscriptstyle \mathrm{-}} + 4U_{\rm NN}\sum_{k\neq i}\frac{m_k}{|r_{ik}|^3} - \frac{4J^2}{U}\sum_{\langle k \rangle_i}m_k,\\
E_{\scriptscriptstyle \mathrm{HP}}^i(J) &= 2\mu_{\scriptscriptstyle \mathrm{-}} - 4U_{\rm NN}\sum_{k\neq i}\frac{m_k}{|r_{ik}|^3} + \frac{4J^2}{U}\sum_{\langle k \rangle_i}m_k.
\end{split}
\end{equation}
Note that in the last expressions, there is no explicit dependence on the chemical potential $\mu_{\scriptscriptstyle \mathrm{+}}$.
This is because by adding or removing one composite, we remain within the subspace at filling factor $\nu$, and therefore the contribution of  
$\mu_{\scriptscriptstyle \mathrm{+}}$ vanishes in the calculation of the excitations (\ref{EQ:PHE_CSS}).
By using a perturbative mean-field method, we can calculate the order parameters $\psi_i=\langle \hat{c}_i \rangle$
for $\ket{\alpha}$, which satisfy the equations
\begin{eqnarray}
 \psi_i &=& \frac{2J^2}{U}\Big[ \frac{\nu(\nu+1) - m_i(m_i+1)}{E_{\scriptscriptstyle \mathrm{PH}}^i(J)} \nonumber \\
 && \; + \; \frac{\nu(\nu+1) - m_i(m_i-1)}{E_{\scriptscriptstyle \mathrm{HP}}^i(J)} \Big]\bar{\psi_i},\label{Eqn:MFCSS}
\end{eqnarray}
where $\bar{\psi_i} = \sum_{\langle j \rangle_i} \psi_j$. With Eqs. (\ref{Eqn:MFCSS}) one can calculate 
the mean-field lobes of any distribution of atoms in the lattice $\ket{\alpha}$, provided that the elementary 
excitations (\ref{EQ:PHE_CSS}), 
are positive in some parameters range. For every site $i$ of the lattice, one can adequately flip the direction of a dipole, and 
depending on $m_i$ either one of the following conditions apply
\begin{equation}
\label{EQ:ConditionCSS}
\begin{split}
\mu_{\scriptscriptstyle \mathrm{-}} &< 2U_{\rm NN}\sum_{k\neq i}\frac{m_k}{|r_{ij}|^3} \\
\mu_{\scriptscriptstyle \mathrm{-}} &> 2U_{\rm NN}\sum_{j\neq i}\frac{m_j}{|r_{ij}|^3}.
\end{split}
\end{equation}
For example, suppose the site $i$ is occupied only by 
one particle of the species $a$, i.e $m_i = 1/2$, then for this site the conditions (\ref{EQ:ConditionCSS}) reduce to 
$\mu_{\scriptscriptstyle \mathrm{-}} > 2U_{\rm NN}\sum_{j\neq i}m_j/|r_{ij}|^3 $, since the only possible excitation at this site corresponds to remove a composite. 
Conditions (\ref{EQ:ConditionCSS}) are necessary for the existence of an insulating lobe and provide its boundaries at $J=0$. For each site of the lattice one has such a condition (\ref{EQ:ConditionCSS}), and an Eq. (\ref{Eqn:MFCSS}).
The latter constitutes a set of coupled equations, which can be written in the matrix form $\mathcal{M}(\mu_{\scriptscriptstyle \mathrm{-}},U,U_{\rm NN},J) \cdot \vec{\psi} = 0$, with 
$ \vec{\psi} \equiv (\cdots \psi_i \cdots )^T$ being the vector of the order parameters at each site of the lattice. 
For every $\mu_{\scriptscriptstyle \mathrm{-}}$, a non-trivial solution is provided by the smallest $J$ for which 
${\rm det} [\mathcal{M}(\mu_{\scriptscriptstyle \mathrm{-}},U,U_{\rm NN},J)] = 0$, that is the insulating lobe of the $\ket{\alpha}$ configuration in the 
$J$ vs. $\mu_{\scriptscriptstyle \mathrm{-}}$ plane.
In Fig. \ref{FIG:GroundStateCSS} we plot the ground state insulating lobes calculated in this way for $\nu=1/2$ (left) and $\nu=1$ (right).
For all filling factors $\nu$, we find an AM ground state $(\nu,M=0)$, which presents a spatial distribution of alternating 
sites occupied by particles of species $a$ and $b$ resembling a checkerboard structure. Remarkably, the larger the $\nu$, and the more stable is
the AM ordering with respect to flipping the direction of a dipole. Indeed, using the inequalities (\ref{EQ:ConditionCSS}), it is not difficult to calculate the boundaries at $J=0$ of such a checkerboard structure, which, for a fixed $\nu$ are given by 
$-2z\nu \ell \, U_{\rm NN} < \mu_{\scriptscriptstyle \mathrm{-}} < 2z\nu \ell \, U_{\rm NN} $, where $\ell = 1-2^{-3/2}-2^{-3}+2\times5^{-3/2}$, and $z = \sum_{\langle j \rangle_i} 1$ is the coordination number (here $z=4$). 
Instead the tip of the AM lobes is found to be $\nu$-independent, and given by $J/U = \sqrt{\ell \,U_{\rm NN}/2U}$
at $\mu_{\scriptscriptstyle \mathrm{-}} = 0$.
By increasing the absolute value of $\mu_{\scriptscriptstyle \mathrm{-}}$ we find RM ground states with rational values of the average magnetization, corresponding to 
$M = (\pm 2\nu, \; \pm 4\nu, \; \pm 6\nu)/8$. The exact fractional values of $M$ in the ground state, depend on the cutoff range in the dipolar interactions, and on the size of the lattice. We have used a $4\times 4$ elementary cell with periodic boundary conditions, and dipolar interaction range cut at the 4th nearest neighbor. By considering more neighbors in the interactions, and larger lattices, we expect to find RM states appearing at all rational $M$, asymptotically approaching a Devil's staircase as recently shown in \cite{BB:Barbara, BB:Cooper}. Finally we find a FM ground state $(\nu,M=\pm \nu)$, in which only particles of one type are present.

It is worth noticing that the insulating lobes calculated in this way, do not contain any dependence on $\mu_{\scriptscriptstyle \mathrm{+}}$,
which does not enter into Eqs. (\ref{Eqn:MFCSS}) as previously discussed.
Therefore, for any given value of $\mu_{\scriptscriptstyle \mathrm{+}}$, in order to obtain the ground state phase diagram one has to compare the energies of the ground state configurations at different $\nu$. Using the effective Hamiltonian (\ref{Heff0_num}), for any value of $\mu_{\scriptscriptstyle \mathrm{+}}$, $J$, and $\mu_{\scriptscriptstyle \mathrm{-}}$, we compare the energies of the ground state configurations for different $\nu$, and select the state with the smaller energy. In this way we have obtained the phase diagram at $J=0$ of Fig. \ref{FIG:GSJ0}.

\subsection{Counterflow superfluid/supersolid}
\begin{center}
\begin{figure}[t]
\begin{center}
\includegraphics[width=1\linewidth]{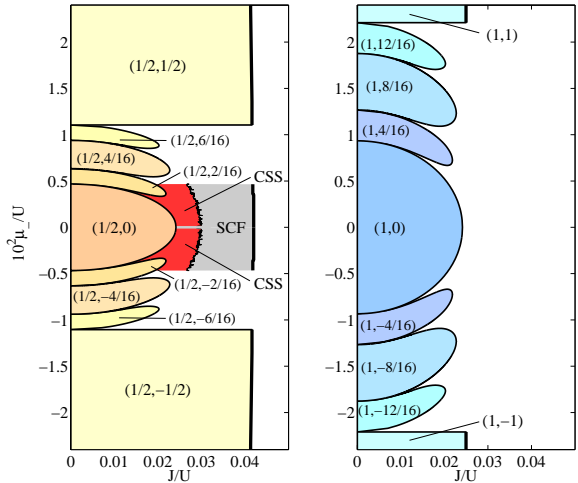}
\end{center}
\caption{Ground state of a $4\times4$ square lattice satisfying periodic boundary conditions, for $\nu=1/2$ at 
$\mu_{\scriptscriptstyle \mathrm{+}} = 0.5U$ (left), and $\nu=1$ at $\mu_{\scriptscriptstyle \mathrm{+}} = 1.4U$ (right),
and $U_{\rm NN} = U/600$. The text in parentheses $(\nu,M)$ indicate the filling factor $\nu$, and the average magnetization $M$, respectively.}
\label{FIG:GroundStateCSS}
\end{figure}
\end{center}

In the low-energy subspace at constant $\nu$, the Gutzwiller Ansatz on the 
wave function of the system reads
\begin{equation}
\label{EQ:NuConstant}
\ket{\Phi} = \prod_i \sum_{\rm m=-\nu}^{\rm \nu} f_{\rm \nu,m}^{(i)} \ket{\nu,m}_i,
\end{equation}
where we allow the Gutzwiller amplitudes to depend on time, i.e. $f_{\rm \nu,m}^{(i)}(t)$.
We obtain the equations of motion for the amplitudes by minimizing the action of the system, given by $S = \int \ud t \mathcal{L}$, with respect to the variational parameters $f_{\rm \nu,m}^{(i)}(t)$ and their complex conjugates $f_{\rm \nu,m}^{*(i)}(t)$, where 
\begin{equation}
\mathcal{L} = i\hbar\frac{\braket{\Phi}{\dot{\Phi}} - \braket{\dot{\Phi}}{\Phi}}{2} - \bra{\Phi} \hat{H}_{\rm eff}^{(0)}\ket{\Phi},
\end{equation}
is the Lagrangian of the system in the quantum state $\ket{\Phi}$ \cite{BB:Perez}.
Therefore setting to zero the variation of the action with respect to $f_{\rm \nu,m}^{*(i)}$ leads to the equations
\begin{eqnarray}
i\hbar \frac{\ud}{\ud t} f_{\rm \nu,m}^{(i)} &=& 
\Big[-2\mu_{\scriptscriptstyle \mathrm{-}} - \frac{4J^2}{U}\sum_{\langle j \rangle_i} \langle \hat{m}_j\rangle \nonumber \\&+& 4U_{NN}\sum_{j\neq i}  \frac{\langle\hat{m}_j \rangle}{|r_{ij}|^3} \Big] m f_{\rm \nu,m}^{(i)} \nonumber \\
&-& \frac{2J^2}{U} \Big[\bar{\psi}_i \sqrt{\nu(\nu+1) - m(m-1)} \; f_{\rm \nu,m-1}^{(i)} \nonumber \\
&+& \bar{\psi}_i^*\sqrt{\nu(\nu+1) - m(m+1)} \; f_{\rm \nu,m+1}^{(i)} \Big]  \label{EQ:FDynamicsEffCSS},
\end{eqnarray}
where $\langle \hat{m}_i \rangle= \sum_{\rm m=-\nu}^{\rm \nu} m |f_{\rm \nu,m}^{(i)}|^2$, the fields $\bar{\psi}_i = \sum_{\langle j \rangle_i} \psi_j$, 
$\sum_{\langle j \rangle_i}  \langle \hat{m}_j\rangle$, and $\sum_{j\neq i}  \langle \hat{m}_j\rangle/|r_{ij}|^3$ have to be calculated in a self consistent way, and the
order parameter $\psi_i = \bra{\Phi} \hat{c}_i \ket{\Phi}$ is given by 
\begin{equation}
\psi_i = \sum_{\rm m=-\nu}^{\rm \nu} \sqrt{\nu(\nu+1) - m(m+1)} \; f_{\rm \nu,m}^{*(i)} f_{\rm \nu,m+1}^{(i)}.
\end{equation}
We solve Eqs. (\ref{EQ:FDynamicsEffCSS}) in imaginary time $\tau=it$, which due to dissipation is supposed to converge to the ground state. 
In Figure \ref{FIG:GroundStateCSS} we show the ground state phase diagram of the system for $\nu = 1/2$ (left) and $\nu=1$ (right), computed in this way 
for $U_{\rm NN} = U/600$. The phase diagram is symmetric with respect to the $\mu_{\scriptscriptstyle \mathrm{-}} = 0$ axis.
This is because the description of the system is identical under the interchange of the species $a$ with the species $b$, and vice versa.
For $\nu=1/2$, in the region immediately outside the insulating AM lobe, depending on the values of $J$ 
and $\mu_{\scriptscriptstyle \mathrm{-}}$ we find either SCF or CSS. 
Note that we don't find any evidence of CSS at 
$\mu_{\scriptscriptstyle \mathrm{-}} = 0$, which indicates that a finite imbalance between the two
species is a necessary condition for the system in order to sustain CSS. 

Let us underline that for $\nu=1/2$, our effective Hamiltonian (\ref{Heff0_num}) can be mapped onto a hard-core dipolar Bose-Hubbard Hamiltonian,
provided that we neglect a constant factor proportional to $\nu$, and rescale the first nearest neighbor interaction by the small quantity $-2J^2/U$. 
Contrary to mean-field predictions, Monte Carlo studies of the hard-core Bose-Hubbard Hamiltonian on square lattices, with interactions extended to first and second nearest neighbors, have shown that no
supersolid phase is obtained by doping the checkerboard solid (AM ordering) \cite{BB:Batrouni,BB:Pollet}. However, it was recently demonstrated
by Monte Carlo simulations that considering an infinite range in the dipolar interactions stabilizes the supersolid phase, obtained by doping the checkerboard solid (AM)
either with particles or holes \cite{BB:Barbara}, in agreement with mean-field predictions. In our treatment we are approaching this limit, and we therefore expect the CSS phase to be stabilized by the long-range
interactions. Furthermore, because $J>0$, the rescaling of the first nearest neighbor interaction by the small quantity $-2J^2/U$ in the mapping process, should enforce further this effect. Although our treatment is approaching infinity in the interaction range, it would be important to verify our predictions with first-principle quantum Monte Carlo simulations.
To get reliable results, one should combine the Gutzwiller
predictions with an estimate of the limits of validity of $\hat{H}_{\rm
eff}^{(0)}$, beyond which the subspace of constant $\nu$ looses its
meaning. Before starting the discussion on the validity
of the subspace, let us explain how we define the
dominant classical configurations of a given state $\ket{\Phi}$. 
It is not difficult to see that Eq. (\ref{EQ:NuConstant}) can be equivalently written as
\begin{equation}
\label{EQ:NuConstant02}
\ket{\Phi} = \sum_{\{\vec{m}\}} g_{\vec{m}}\prod_i \ket{\nu,m_i}_i,
\end{equation}
where $\vec{m} = (m_1,...,m_i,..., m_{\scriptscriptstyle N_{\scriptscriptstyle \mathrm{S}}})$ is a collection of the 
indices $m$ at each site, and we have introduced the notation $g_{\vec{m}} = \prod_i f_{\rm \nu, m_i}^{(i)}$. 
The advantage of writing
the Gutzwiller state $\ket{\Phi}$ in the form (\ref{EQ:NuConstant02}), lays in the product over single-site Fock
states $\ket{\alpha} = \prod_i \ket{\nu,m_i}_i$, which is nothing but a classical distribution of atoms in the lattice. 
Therefore we can rewrite Eq. (\ref{EQ:NuConstant02}) as
\begin{equation}
\label{EQ:GWAlphas}
 \ket{\Phi} = \sum_{\{\alpha\}} g_{\rm \alpha}\ket{\alpha}.
\end{equation}
For each point of the phase diagram, from the ground
state Gutzwiller wavefunction, we define the dominant classical
configurations with the criteria
$|g_{\vec{m}}| = |\prod_i f_{\rm \nu,m_i}^{(i)}|>(0.02)^4$, and we require $|f_{\rm \nu,m_i}^{(i)}|^2>0.05$,
implying that each of the contributing $f_{\rm \nu,m_i}^{(i)}$
should also be sufficiently large.
For each of these configurations, we calculate the lobe with respect to single
particle-hole excitations \footnote{We have checked that the validity region is not
strongly modified upon small changes in these conditions.}. If the system at this given point of
the phase diagram turns out to be stable against all dominant
single particle-hole excitations (in other words, if this point is
inside all selected single particle-hole lobes), $\hat{H}_{\rm
eff}^{(0)}$ is considered valid. This procedure is
shown for $\mu_{\scriptscriptstyle \mathrm{+}} = 0.4U$, $J=0.035U$ and $\mu_{\scriptscriptstyle \mathrm{-}} = 0$ in Fig.~\ref{FIG:StabilityCSS}, 
and gives the thick black lines of Fig. (\ref{FIG:GroundStateCSS}). On the right hand side of 
these black lines, the low energy subspace is not well defined and therefore the effective 
Hamiltonian looses its meaning, leaving the description of the system to the
domain of single-particle single-hole excitation theory that predicts 
SF and SS phases for each component separately.

\begin{center}
\begin{figure}[t]
\begin{center}
\includegraphics[width=1\linewidth,keepaspectratio=true]{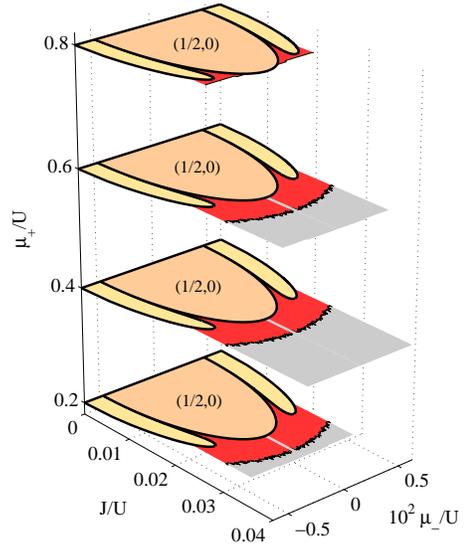}
\end{center}
\caption{Slices of the 3D ground state of a $4\times4$ square lattice satisfying periodic boundary conditions, for $\nu=1/2$ and $U_{\rm NN} = U/600$.}
\label{FIG:GroundState3D}
\end{figure}
\end{center}
We have already mentioned, that the boundaries of the lobes calculated with the effective Hamiltonian do not show any dependence 
on the chemical potential $\mu_{\scriptscriptstyle \mathrm{+}}$, which does not give any contribution in the expression of the low-lying excitations (\ref{EQ:PHE_CSS}). 
This is not true in the case of the single-particle-hole insulating lobes,
since adding or removing a single particle results in a change of both $\mu_{\scriptscriptstyle \mathrm{+}}$, and $\mu_{\scriptscriptstyle \mathrm{-}}$.
This makes the process of estimating the limits of validity of $\hat{H}_{\rm eff}^{(0)}$ more complicated and leads to a 3D phase diagram in $J$,
$\mu_{\scriptscriptstyle \mathrm{+}}$, and $\mu_{\scriptscriptstyle \mathrm{-}}$ highly non-trivial.
In Fig. \ref{FIG:GroundState3D} we present slices of the 3D phase diagram, calculated at constant values of $\mu_{\scriptscriptstyle \mathrm{+}}$, for $\nu = 1/2$ only. From Fig. \ref{FIG:GroundState3D}, it is evident that the insulating magnetic phases persist for a wide range of $\mu_{\scriptscriptstyle \mathrm{+}}$ values. Instead, the SCF and the CSS phases are more affected by the limits of validity 
of $\hat{H}_{\rm eff}^{(0)}$, which may imply high degrees of control in order to experimentally observe this quantum phases.

Finally, it is useful to estimate the critical temperatures at which we expect these quantum phases to be experimentally observable. Bose-Einstein condensation in alkali atoms occurs at temperatures of the order $~\sim 100$ nK. Assuming that condensation of OH molecules occurs at similar temperatures we expect the SCF to be observed at $\sim 100$ nK. For the AM insulating phases, the gap at $J=0$ is of the order $ \sim 2\nu\times 10^{-2} U$, and the critical temperature is expected to be of the order of the standard insulator-superfluid transition temperature, i.e. $10$ to $50$ nK \cite{BB:Maciek01}. Therefore, the CSS phase is expected to be observed somewhere in between $50$ to $100$ nK.
The lowest measured temperature of ultracold atoms in optical lattices is $\sim 1$ nK \cite{BB:Ketterle}, where superexchange interactions, of the order $\sim J^2/U$, are dominant. Recently, a two-component hard-core Bose-Hubbard Hamiltonian was investigated with Monte Carlo techniques \cite{BB:Barbara02}, and insulating AM as well as
SCF phases where observed. In \cite{BB:Barbara02}, the largest critical temperature observed for the AM phase is $T_c \sim 0.12 J$, at $J/U=0.0125$, and $T_c \sim 0.104 J$
at $U/J = 13$ for the SCF phase.

\section{Conclusions}
In conclusion, in this work we have studied a gas of dipolar Bosons confined in a two-dimensional optical lattice, where the dipoles are free to 
point in both directions perpendicular to the lattice plane. We have found regions of parameters where the ground state presents insulating phases with antiferromagnetic or ferromagnetic ordering, as well as with rational values of the average magnetization. Our mean-field calculations predict the existence of a novel {\it counterflow super solid} quantum phase, which presents a spatial modulation of the density coexisting with the presence of counterflow. Our work in a sense is the first step  in the studies of dipolar gases with non-polarized  dipoles, and is relevant for experimental studies of the ultracold OH molecules. We expect that the methods and ideas developed here will turn out to be useful also for the study of more complex systems, such as excitons in indirect quantum well structures, or completely unpolarized dipolar gases. 

\subsection{Acknowledgements}
The authors thank B.~Lev, T.~Roscilde, and G.~Shlyapnikov 
for interesting discussions.
We acknowledge support of Spanish MEC and MINCIN (TOQATA FIS2008-
00784, QOIT), EU projects SCALA, AQUTE and NAMEQUAM,
and ERC grant QUAGATUA. M.L. thanks also Alexander von Humboldt foundation for support.


\begin{thebibliography}{99}
\bibitem{BB:Foote01}
In standard experiments with $^{52}$Cr, which features a magnetic moment of $\mu = 6\mu_{\scriptscriptstyle \mathrm{B}}$,
this ratio is given by $U_{\rm NN} \simeq U/400$, for an optical lattice depth of $20E_{\rm R}$, where $E_{\rm R}$ is the recoil 
energy at $\lambda = 500$nm.
\end{thebibliography}
\end{document}